\documentclass[aps,prx,twocolumn,nofootinbib,groupedaddress,superscriptaddress,longbibliography]{revtex4-2}

\usepackage{amssymb}
\usepackage{multirow}
\usepackage{graphicx}
\usepackage{amsmath}
\usepackage{color}
\usepackage{mathrsfs}
\usepackage{float}
\usepackage{indentfirst}
\usepackage{mathrsfs}
\usepackage{float}
\usepackage{indentfirst}
\usepackage{textcomp}
\usepackage{comment}
\usepackage{mathtools}
\usepackage{natbib,hyperref}
\usepackage{soul}

\begin{document}
\title{Augmenting Finite Temperature Tensor Network with Clifford Circuits}

\author{Xiangjian Qian}
\affiliation{Key Laboratory of Artificial Structures and Quantum Control (Ministry of Education),  School of Physics and Astronomy, Shanghai Jiao Tong University, Shanghai 200240, China}

\author{Jiale Huang}
\affiliation{Key Laboratory of Artificial Structures and Quantum Control (Ministry of Education),  School of Physics and Astronomy, Shanghai Jiao Tong University, Shanghai 200240, China}

\author{Mingpu Qin} \thanks{qinmingpu@sjtu.edu.cn}
\affiliation{Key Laboratory of Artificial Structures and Quantum Control (Ministry of Education),  School of Physics and Astronomy, Shanghai Jiao Tong University, Shanghai 200240, China}

\affiliation{Hefei National Laboratory, Hefei 230088, China}

\date{\today}


\begin{abstract}
    Recent studies have highlighted the combination of tensor network methods and the stabilizer formalism as a very effective framework for simulating quantum many-body systems, encompassing areas from ground state to time evolution simulations. In these approaches, the entanglement associated with stabilizers is transferred to Clifford circuits, which can be efficiently managed due to the Gottesman-Knill theorem. Consequently, only the non-stabilizerness entanglement needs to be handled, thereby reducing the computational resources required for accurate simulations of quantum many-body systems in tensor network related methods. In this work, we adapt this paradigm for finite temperature simulations in the framework of Time-Dependent Variational Principle, in which imaginary time evolution is performed using the purification scheme. Our numerical results on the one-dimensional Heisenberg model and the two-dimensional $J_1-J_2$
    Heisenberg model demonstrate that Clifford circuits can significantly improve the efficiency and accuracy of finite temperature simulations for quantum many-body systems. This improvement not only provides a useful tool for calculating finite temperature properties of quantum many-body systems, but also paves the way for further advancements in boosting the finite temperature tensor network calculations with Clifford circuits and other quantum circuits.
\end{abstract}

\maketitle
{\em Introduction --}
Solving strongly correlated quantum many-body systems presents significant challenges due to the exponential growth of the Hilbert space and intricate quantum correlations. Existing numerical methods, such as Density Matrix Renormalization Group (DMRG) and its descendants \cite{PhysRevLett.69.2863,RevModPhys.77.259,10.5555/2011832.2011833,SCHOLLWOCK201196,RevModPhys.93.045003,1992CMaPh.144..443F,tao_book}, have demonstrated their efficacy in solving one-dimensional (1D) systems. However, they face challenges in higher dimensions due to the constraints imposed by entanglement entropy within these ansatzes. Attempts \cite{Qian_2023} have been made to increase the entanglement encoded while keeping the low cost in Matrix Product States (MPS) \cite{PhysRevLett.75.3537}, which is the underlying wave-function ansatz of DMRG. We also notice that 2D ansatzes \cite{RevModPhys.93.045003,tao_book,2004cond.mat..7066V,PhysRevLett.99.220405,PhysRevLett.102.180406,PhysRevX.4.011025,PhysRevB.81.165104} exists by generalizing MPS to 2D but with higher cost.

It is known that there exist classical simulatable quantum circuits which can support large entanglement. For example, Clifford circuits which are composed exclusively of Clifford gates (Hadamard, S, and Controlled-NOT gates) \cite{Nielsen_Chuang_2010} can be efficiently simulated on a classical computer according to the Gottesman-Knill theorem \cite{gottesman1997stabilizer,PhysRevA.70.052328,PhysRevA.73.022334}. The states that can be prepared using these gates are known as stabilizer states \cite{lami2024learning,PhysRevA.70.052328,PhysRevA.73.022334,sun2024stabilizer,gottesman1997stabilizer,PhysRevLett.128.050402}. These states can exhibit significant entanglement while remaining classically simulatable \cite{PhysRevX.7.031016}. However, quantum circuits consisting solely of Clifford gates are not universal for quantum computing. To achieve universality, additional quantum gates, often referred as non-stabilizerness or “magic” gates, must be incorporated \cite{PhysRevLett.128.050402,lami2024learning,tarabunga2024nonstabilizerness,PhysRevLett.131.180401,frau2024nonstabilizerness,PhysRevLett.112.240501,PRXQuantum.3.020333,PhysRevA.71.022316,10.21468/SciPostPhys.16.2.043,masotllima2024stabilizertensornetworksuniversal,lami2024quantum}. 

It is tempting to combine MPS and Clifford circuits \cite{doi:10.1021/acs.jctc.3c00228,lami2024quantum,masotllima2024stabilizer}, giving their complementary properties. But how to optimize the ansatz is the key issue. In \cite{Qian_2023}, MPS are augmented with general disentanglers which makes the computational cost increase with the layer of disentanglers. Clifford circuits can be viewed as special disentanglers that persevere Pauli string under conjugation, making the augment of MPS with infinite layers of Clifford circuits possible while keeping the cost under control. Recently, we developed a seamless integration of Clifford circuits in two-site DMRG \cite{qian2024augmentingdensitymatrixrenormalization}, optimizing the structure of the wave-function ansatz and the local tensors and disentanglers simultaneously with negligible overhead.  
Numerical results show that the Clifford Circuits Augmented Matrix Product States (CA-MPS) method \cite{qian2024augmentingdensitymatrixrenormalization} can improve the simulation accuracy significantly. This innovative approach leverages the efficient classical simulation of stabilizer states alongside the efficient classical simulation of tensor network states, particularly MPS. Subsequent research has further highlighted the effectiveness of integrating Clifford circuits to enhance tensor network simulations
\cite{huang2024nonstabilizernessentanglemententropymeasure,fux2024disentanglingunitarydynamicsclassically,qian2024cliffordcircuitsaugmentedtimedependent,Antonio0701,PhysRevLett.133.150604,doi:10.1021/acs.jctc.3c00228}. This framework has also been extended to improve the simulation of time evolution \cite{qian2024cliffordcircuitsaugmentedtimedependent,Antonio0701,PhysRevLett.133.150604}. This development also expands our understanding on classically simulatable quantum states \cite{huang2024nonstabilizernessentanglemententropymeasure}. 

\begin{figure*}[t]
    \includegraphics[width=190mm]{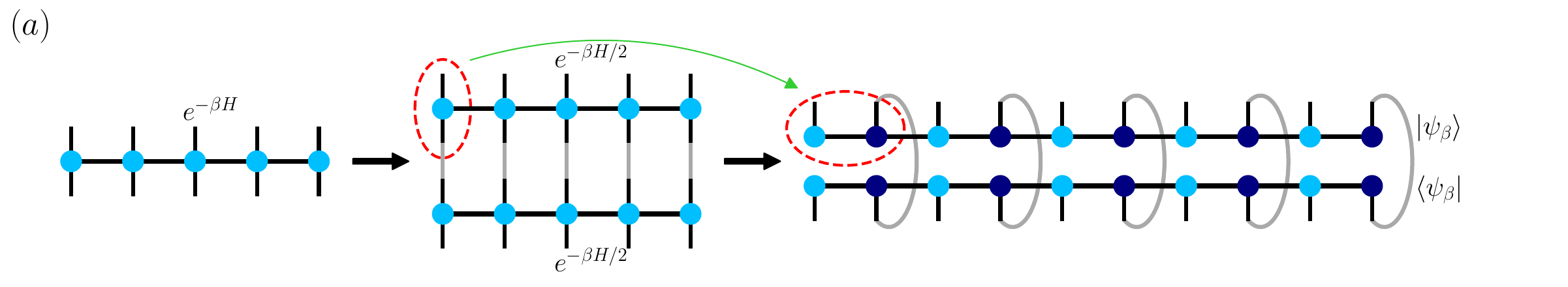}
    \includegraphics[width=190mm]{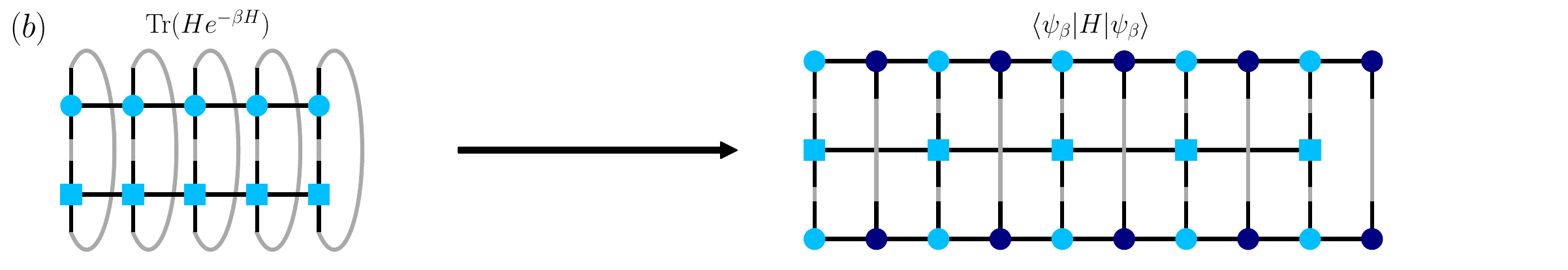}
       \caption{Transforming the thermal canonical density matrix for finite temperature TDVP simulations. (a) An auxiliary Hilbert space Q (denoted as the navy dots) is introduced on the even sites to purify the thermal canonical density matrix. Then, the thermal canonical density matrix can be expressed as $\rho_{\beta}=e^{-\beta H}=\text{Tr}_Q |\psi_{\beta}\rangle\langle \psi_{\beta}|$, with $|\psi_{\beta}\rangle=e^{-\beta H/2}|\psi_{0}\rangle$ ($|\psi_{0}\rangle$ is the thermal canonical density matrix at infinite temperature $(\beta=0)$). (b) The purification facilitates the efficient calculation of physical observables, e.g., the internal energy $\text{Tr}(He^{-\beta H})$ by evaluating the expectation value $\langle \psi_{\beta}|H|\psi_{\beta}\rangle$, where $H$ acts solely on the odd sites.} 
       \label{Choi}
\end{figure*}

Other than ground state and time evolution simulations, finite temperature simulations provide a comprehensive understanding of quantum many-body systems from another angle. They allow for the calculation of quantities such as heat capacity, which facilitates the direct comparisons of experimental and numerical results.
Simulating thermal many-body states initially seems more challenging than ground state simulations due to the exponentially large number of excited states in the energy spectrum, which may make simulation on classical computers difficult. However, the ensemble density operator $e^{-\beta H}$ (with $\beta= 1/T$ being the inverse temperature) can be efficiently represented and manipulated using thermal tensor network method \cite{RevModPhys.82.277,PhysRevA.78.022103,PhysRevB.56.5061,PhysRevB.58.9142,PhysRevLett.102.190601,Stoudenmire_2010,PhysRevX.8.031082}.
The Matrix Product Operator (MPO) serves as a particularly apt thermal tensor network for modeling 1D quantum systems at finite temperatures.
Nevertheless, the required (MPO) bond dimension scales as $D \sim e^{\beta}$ \cite{PhysRevA.78.022103,PhysRevB.73.085115,PhysRevX.8.031082,RevModPhys.82.277} for critical systems, which poses a significant challenge for accessing low-temperature properties. In this work, we aim to extend the CA-MPS framework to simulate quantum many-body systems at finite temperatures, seeking to advance current methodologies for finite temperature simulations.

In this work, we primarily focus on employing the Time-Dependent Variational Principle (TDVP) \cite{PhysRevLett.107.070601,PhysRevB.94.165116} for finite temperature simulations. The similarities of TDVP with DMRG enable the straightforward integration of Clifford circuits \cite{qian2024cliffordcircuitsaugmentedtimedependent}. Additionally, it is intriguing to explore the potential integration of Clifford circuits into other finite temperature simulation methods \cite{PhysRevB.56.5061,PhysRevB.58.9142,PhysRevB.72.220401,PhysRevLett.102.190601,Stoudenmire_2010,PhysRevX.8.031082}.

In the rest of this work, we discuss the details of the implementation of finite temperature Clifford Augmented Time-Dependent Variational Principle (CA-TDVP) method and evaluate its performance in both 1D and 2D systems. Our results demonstrate the method's capability to significantly improve the efficiency of finite temperature simulations.

\begin{figure*}[t]
    \includegraphics[width=80mm]{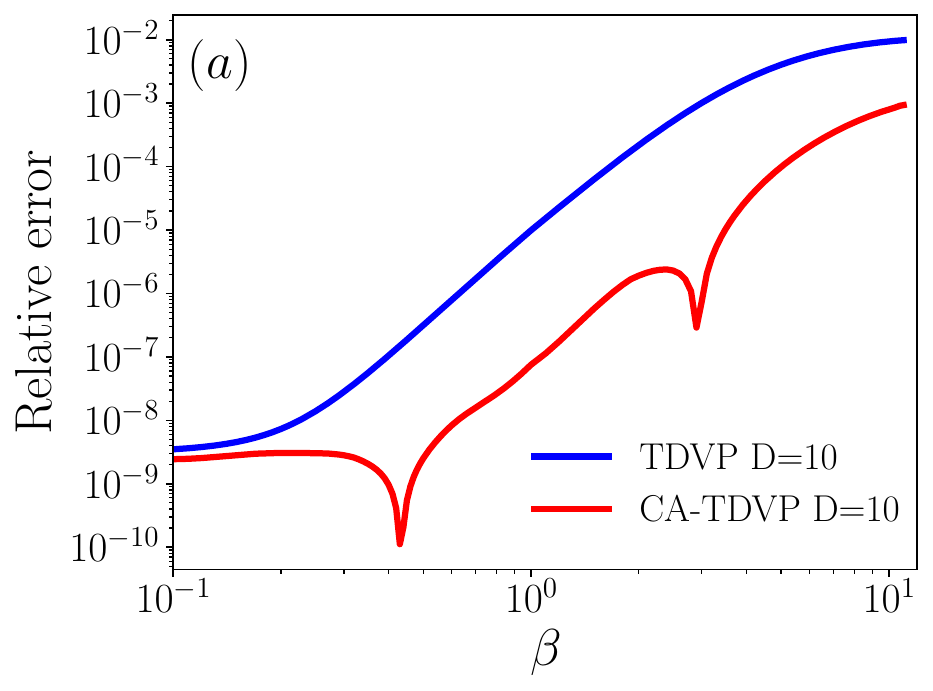}
    \includegraphics[width=80mm]{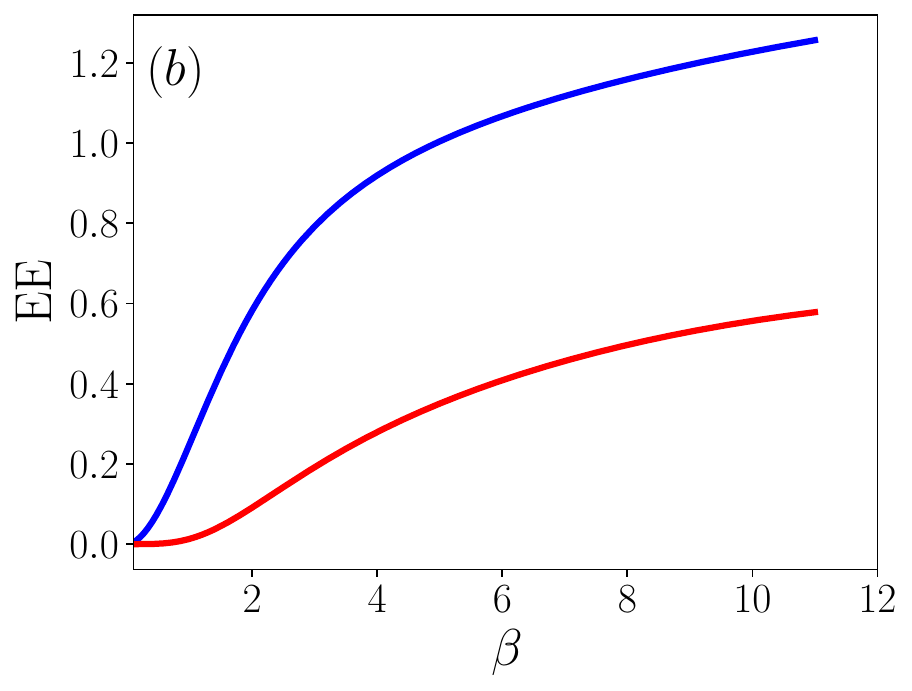}
       \caption{
       The simulation results for the 1D Heisenberg model. The system is a 1D chain with a length of $N=18$ under open boundary conditions. (a) The relative error of the internal energy as a function of inverse temperature $\beta$ with the results calculated by Exact Diagonalization (ED) as a reference. (b) the entanglement entropy (EE) at the center bond in the MPS part as a function of inverse temperature $\beta$.} 
       \label{1D_Hei}
\end{figure*}

\begin{figure*}[t]
    \includegraphics[width=80mm]{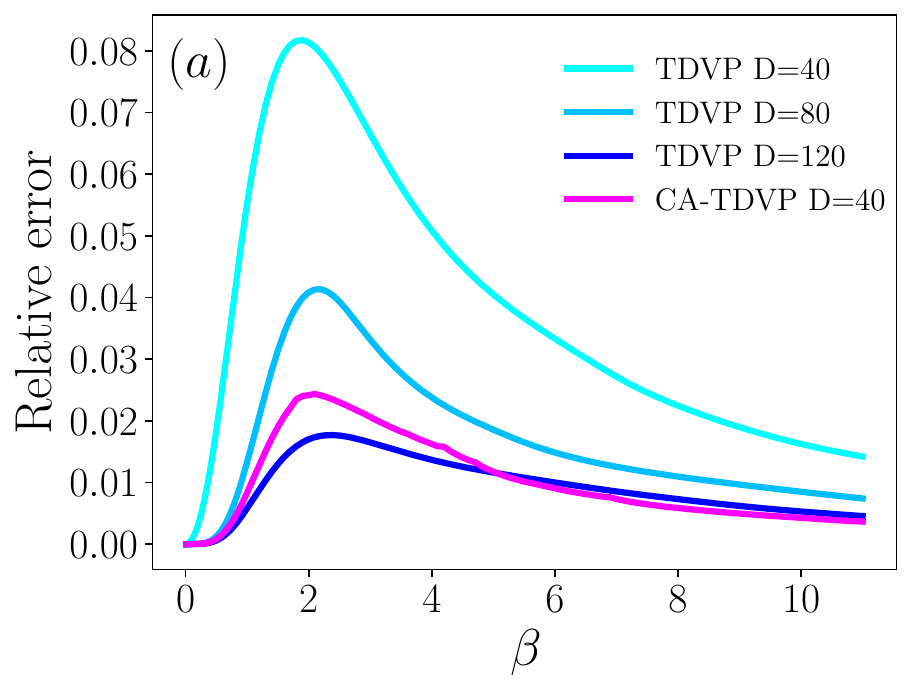}
    \includegraphics[width=80mm]{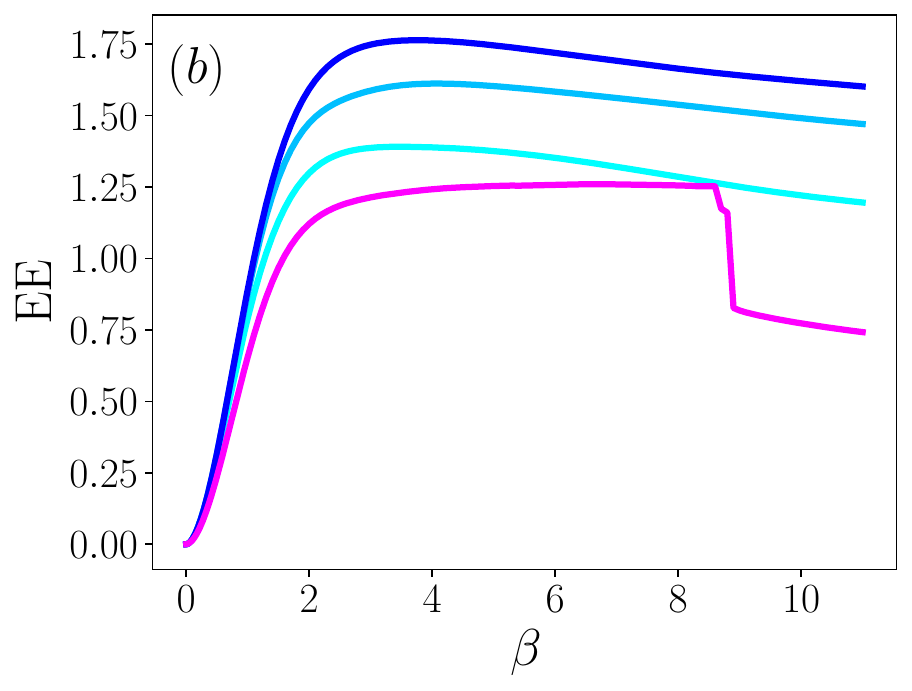}
       \caption{
        The simulation results for the 2D Heisenberg $J_1-J_2$ model at $J_2=0$. The system is with size $4\times 4$ under open boundary conditions. (a) The relative error of the internal energy as a function of inverse temperature $\beta$ with the results calculated by Exact Diagonalization (ED) as a reference. (b) the entanglement entropy (EE) at the center bond in the MPS part as a function of inverse temperature $\beta$.}
       \label{2D_Hei}
\end{figure*}

\begin{figure*}[t]
    \includegraphics[width=80mm]{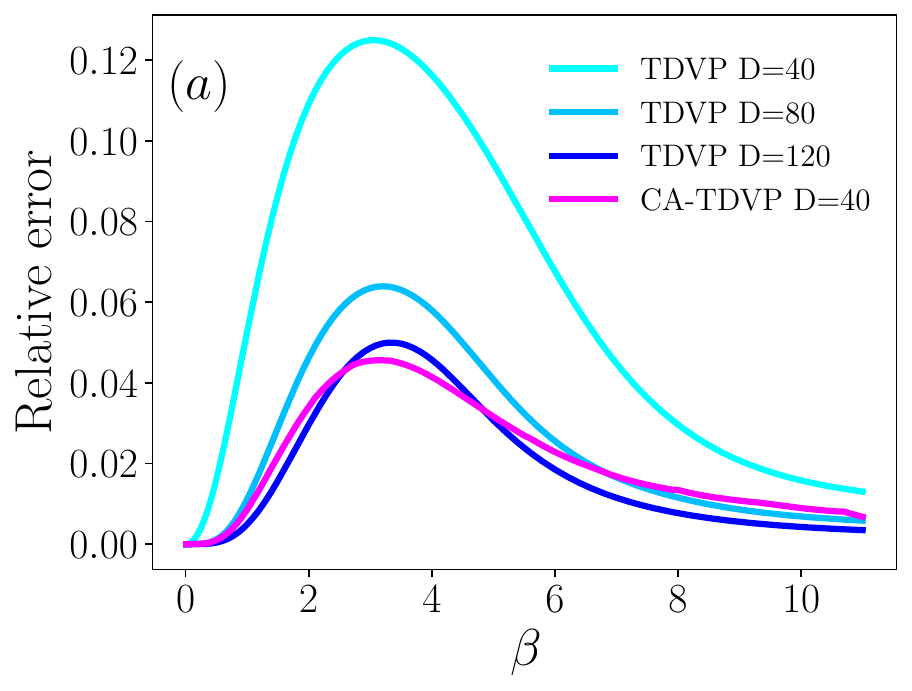}
    \includegraphics[width=80mm]{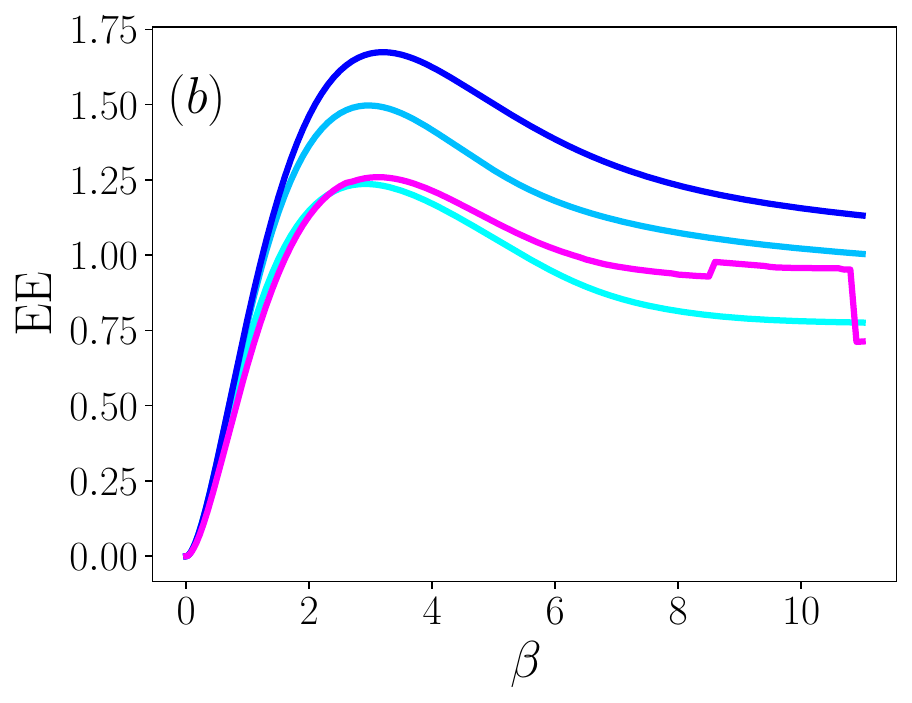}
       \caption{
       The simulation results for the 2D Heisenberg $J_1-J_2$ model at $J_2=0.5$. The system is with size $4\times 4$ under open boundary conditions. (a) The relative error of the internal energy as a function of inverse temperature $\beta$ with the results calculated by Exact Diagonalization as a reference. (b) the entanglement entropy (EE) at the center bond in the MPS part as a function of inverse temperature $\beta$.}
       \label{2D_J1J2}
\end{figure*}

{\em CA-TDVP method for finite temperature simulations--}
One scheme to perform finite temperature calculation is to purify the thermal canonical density matrix \cite{Suzuki:1985ypr,SCHOLLWOCK201196}
by introducing an auxiliary Hilbert space $Q$: 
\begin{equation}
    \rho_{\beta}=e^{-\beta H}=\text{Tr}_Q |\psi_{\beta}\rangle\langle \psi_{\beta}|
\end{equation} 
with $|\psi_{\beta}\rangle=e^{-\beta H/2}|\psi_{0}\rangle$ and $|\psi_{0}\rangle$ is presented by a matrix product state
\begin{equation}
    |\psi_{0}\rangle=\sum_{\{\sigma_i\}} \text{Tr}[M_1^{\sigma_1}M_2^{\sigma_2}M_3^{\sigma_3}\cdots M_{2N}^{\sigma_{2N}}]|\sigma_1 \sigma_2 \sigma_3\cdots \sigma_{2N}\rangle
\end{equation}  
where $|\sigma_i\rangle$ denotes the eigenbasis of $S^z_i$, $M^{\uparrow}_{2i-1}=[1, 0],M^{\downarrow}_{2i-1}=[0, 1], M^{\uparrow}_{2i}=[0, 1/\sqrt{2}]^T,M^{\downarrow}_{2i}=[1/\sqrt{2}, 0]$. $|\psi_{0}\rangle$ is the thermal canonical density matrix at infinite temperature $(\beta=0)$, in which the physical and auxiliary spins at positions $2i-1$ and $2i$ are maximally entangled. The auxiliary Hilbert space $Q$ is spanned by states located on even sites, while the physical Hilbert space resides on odd sites and the Hamiltonian ($H$) only acts on the physical Hilbert space. Consequently, the Hamiltonian can be rewritten as
\begin{equation}
    H=\sum_{i=1}^{m}a_i P_i
\end{equation} 
with $P_i=\sigma_1 \otimes I_2 \otimes \sigma_3\cdots \otimes \sigma_{2N-1} \otimes I_{2N}$, $\sigma_i\in \{I, \sigma^x, \sigma^y, \sigma^z\}$, $\sigma^{\alpha}$ is Pauli matrix (assuming that the Hamiltonian has been rewritten as a sum of Pauli strings). After performing such a transformation (an illustration of this transformation is shown in Fig.~\ref{Choi} (a)), we can employ TDVP method with imaginary time to calculate the thermal state at the inverse temperature $\beta$ according to $|\psi_{\beta}\rangle=e^{-\beta H/2}|\psi_{0}\rangle$ \cite{PhysRevLett.107.070601,PhysRevB.94.165116,PhysRevLett.130.226502} and expectation values of any observable $\hat{O} $ with $\langle \hat{O}\rangle_{\beta}=$$\langle\psi_{\beta}|\hat{O}|\psi_{\beta}\rangle$. An example for calculating the internal energy is shown in Fig.~\ref{Choi} (b). Here, we consider the two-site TDVP method. 

The two-site TDVP method \cite{PhysRevLett.107.070601,PhysRevB.94.165116} decomposes  the short inverse temperature evolution $e^{-H\Delta \beta}$ of an MPS into a sweep process analogous to the DMRG algorithm. Specifically, in a left-to-right sweep, we deal with the effective Hamiltonians
\begin{equation}
    \begin{split}
    H_{\text{eff}}&=\sum_{i=1}^{m} a_i A_{i,k-1} \otimes \sigma_{i,k} \otimes \sigma_{i,k+1} \otimes B_{i,k+2}\\
    K_{\text{eff}}&=\sum_{i=1}^{m} a_i A_{i,k} \otimes \sigma_{i,k+1} \otimes B_{i,k+2}\\
    \end{split}
\end{equation}
where $A_{i,k}, B_{i,k}$ are the so-called left and right environment for $P_i$ at site $k$ \cite{SCHOLLWOCK201196}, $\sigma_{i,k}$ is the Pauli matrix of $P_i$ at site $k$ and $a_i$ is the associated interaction strength of $P_i$. 

In two-site TDVP, we update the local tensors $M_k,M_{k+1}$ ($M_i$ is the local MPS tensor at site $i$) through the following steps:
\begin{enumerate}
    \item We first update the local states $|\phi\rangle$ associated with local tensors $M_k,M_{k+1}$ according to $H_{\text{eff}}$: $|\phi\rangle \coloneq e^{-H_{\text{eff}}\Delta \beta/2}|\phi\rangle$.
    \item Next, we perform a Singular Value Decomposition (SVD) on $|\phi\rangle$ to obtain the updated tensors $M_k,M_{\text{k+1}}$ and update the left environment $A_{i,k}$ according to the updated tensor $M_k$. 
    \item Finally, we update the local state $|\psi\rangle$ associated with local tensor $M_{k+1}$ according to $K_{\text{eff}}$: $|\psi\rangle \coloneq e^{K_{\text{eff}}\Delta \beta/2}|\psi\rangle$ (notice that the sign of the exponent in this step is opposite to that in the first step).
\end{enumerate}
By sweeping from $k=1$ to $k=2N$ ($N$ is the system size), the system evolves for $\Delta \beta$: $|\text{MPS} \rangle_{\Delta \beta} \approx e^{-H\Delta \beta/2} |\text{MPS} \rangle$. 

We can enhance the TDVP method by incorporating an additional step with Clifford circuits. Specifically, after calculating $|\phi\rangle \coloneq e^{-H_{\text{eff}}\Delta \beta/2}|\phi\rangle$ in the first step, we incorporate a Clifford circuits $\mathcal{C}$ to reduce the entanglement in $|\phi\rangle$ \cite{qian2024augmentingdensitymatrixrenormalization,qian2024cliffordcircuitsaugmentedtimedependent}:  $|\phi\rangle \coloneq \mathcal{C}|\phi\rangle$. This reduction in entanglement decreases the bond dimension required to represent $|\phi\rangle$ and the truncation loss in the subsequent SVD process in the second step.
Concurrently, we need to perform a transformation to update the original Hamiltonian $H \coloneq \mathcal{C} H \mathcal{C}^\dagger $. We also need to transform any observable to measure, ensuring that physical quantities remain invariant under the action of the Clifford circuit $\mathcal{C}$. Following this modification, the subsequent steps proceed the same as in the standard TDVP method.

The primary advantage of using Clifford circuits $\mathcal{C}$
to disentangle the wave function, as opposed to employing a general unitary matrix \cite{Qian_2023}, lies in the efficiency and simplicity they offer. When a Hamiltonian is conjugated by Clifford circuits, it can be efficiently calculated using the stabilizer formalism \cite{PhysRevA.70.052328}. Furthermore, this method preserves the structural simplicity of the original Hamiltonian, as Clifford circuits transform a Pauli string into another Pauli string, maintaining the form of the sum of Pauli strings and the number of terms of the original Hamiltonian.

{\em Results for 1D Heisenberg model --}
We first test the finite temperature CA-TDVP method on the 1D Heisenberg model under open boundary conditions (OBC). The Hamiltonian of the model is defined as
\begin{equation}
	H = \sum_i^{N-1} S_{i} \cdot S_{i+1}
 \label{Hei_1D_H}
\end{equation}
where $S^i$ is the spin-1/2 operator on site $i$.
Fig.~\ref{1D_Hei} shows the results for a chain with length $N=18$. The inverse temperature step in TDVP is set as $\Delta \beta=0.05$. The relative error of the internal energy compared to the exact results calculated by Exact Diagonalization (ED) \cite{10.21468/SciPostPhys.7.2.020} and the half-chain entanglement entropy (EE) as a function of inverse temperature $\beta$ are plotted. In Fig.~\ref{1D_Hei} (a), we observe that CA-TDVP substantially improves the accuracy by several orders of magnitude compared to pure TDVP simulations, with the same bond dimension $D$. From Fig.~\ref{1D_Hei} (b), we can find the significant reduction of entanglement entropy in CA-TDVP simulations compared to the TDVP results, demonstrating the effectiveness of finite temperature CA-TDVP method.

{\em Results for 2D $J_1-J_2$ Heisenberg model --}
We also test the finite temperature CA-TDVP method on the 2D $J_1-J_2$ Heisenberg model, with Hamiltonian
\begin{equation}
	H = J_1\sum_{\langle i,j \rangle}S_{i} \cdot S_{j}+J_2\sum_{\langle\langle i,j \rangle\rangle}S_{i} \cdot S_{j}
\end{equation}
where $S_i$ is the spin-1/2 operator on site $i$, and the summations are taken over nearest-neighbor ($\langle i,j \rangle$) and next-nearest-neighbor ($\langle \langle i,k \rangle \rangle$) pairs. In our calculations, we fix $J_1=1$. The inverse temperature step in TDVP is set as $\Delta \beta=0.05$. Here, we consider a $4\times 4$ lattice with open boundary conditions. Fig.~\ref{2D_Hei} and Fig.~\ref{2D_J1J2} show the relative error of the internal energy compared to the exact results calculated by ED \cite{10.21468/SciPostPhys.7.2.020} and the half-system entanglement entropy (EE) as a function of inverse temperature $\beta$ for $J_2=0$ and $J_2=0.5$, respectively. We find that CA-TDVP can achieve the same accuracy as TDVP while using only about one-third of the bond dimension as TDVP. For instance, as illustrated in Fig.~\ref{2D_Hei} (a) and Fig.~\ref{2D_J1J2} (a), the results of CA-TDVP with $D=40$ are comparable to those of TDVP with $D=120$, especially at the temperature with largest error. This indicates a significant efficiency improvement of CA-TDVP over TDVP, given that the computational cost for TDVP scales as 
$O(D^3)$ and CA-TDVP has almost the same cost as TDVP. Additionally, the entanglement entropy in CA-TDVP calculations is also significantly reduced as shown in Fig.~\ref{2D_Hei} (b) and Fig.~\ref{2D_J1J2} (b).

{\em Conclusion and Perspective --}
In this work, we introduce a straightforward yet effective enhancement to finite-temperature TDVP by incorporating Clifford circuits in the framework, which can be seamlessly integrated into existing implementations. The finite temperature CA-TDVP results for both 1D Heisenberg model and 2D $J_2-J_2$ Heisenberg model demonstrate a significant improvement over original TDVP simulations. This enhancement offers a clear advantage: it substantially reduces entanglement and increases simulation accuracy while requiring negligible additional computational effort. This advancement enables us to tackle larger and more complex systems that were previously computationally prohibitive.

Our work paves the way for numerous avenues of future research. One promising direction is to explore the integration of Clifford circuits with other finite temperature numerical methods, thereby expanding the toolkit available for finite temperature simulation of quantum many-body systems. We are also exploring quantum circuits which can be integrated with MPS other than Clifford circuits. Furthermore, examining the applicability of this approach across different quantum systems and temperature ranges could provide valuable insights for the physics of quantum many-body systems. Such investigations have the potential to lead to breakthroughs in understanding complex quantum phenomena and further enhance the synergy between numerical simulations and experimental studies. It will be also interesting to apply this method to Fermi systems in the future. 

In summary, the incorporation of Clifford circuits into finite-temperature TDVP bridges the fields of quantum information and quantum many-body physics, representing a significant step in improving the accuracy and efficiency of finite temperature simulations. We anticipate it will play a crucial role in advancing our understanding of finite temperature properties of quantum many-body systems.


\begin{acknowledgments}
The calculation in this work is carried out with TensorKit \cite{foot7}. The computation in this paper were run on the Siyuan-1 cluster supported by the Center for High Performance Computing at Shanghai Jiao Tong University. MQ acknowledges the support from the National Natural Science Foundation of China (Grant No. 12274290), the Innovation Program for Quantum Science and Technology (2021ZD0301902), and the sponsorship from Yangyang Development Fund.
\end{acknowledgments}

\bibliography{main}

\end{document}